\documentclass[
    final            
  ]
  {aipproc}

\layoutstyle{6x9}

\begin{document}

\title{3D calculation of Tucson-Melbourne 3NF effect in triton binding energy}

\classification{21.45.-v, 21.45.Ff, 21.10.Hw}
\keywords      {triton binding energy, three dimensional approach, Bonn-B, Tucson-Melbourne, three-nucleon force}

\author{M.~R.~Hadizadeh}{
  address={Instituto de F\'{\i}sica Te\'{o}rica (IFT), Universidade Estadual Paulista (UNESP), Barra Funda,
01140-070, S\~{a}o Paulo, Brazil.}
}

\author{L. Tomio}{
  address={Instituto de F\'{\i}sica Te\'{o}rica (IFT), Universidade Estadual Paulista (UNESP), Barra Funda,
01140-070, S\~{a}o Paulo, Brazil.}
}

\author{S.~Bayegan}{
  address={Department of Physics, University of Tehran, P.O.Box 14395-547, Tehran, Iran.}
}

\begin{abstract}
As an application of the new realistic three-dimensional (3D) formalism reported recently for three-nucleon (3N) bound states, an attempt is made to study the effect of three-nucleon forces (3NFs) in triton binding energy in a non partial
wave (PW) approach. The spin-isospin dependent 3N Faddeev integral equations with the inclusion of 3NFs, which are formulated as function of vector Jacobi momenta, specifically the magnitudes of the momenta and the angle between them, are solved with Bonn-B and Tucson-Melbourne NN and 3N forces in operator forms which can be incorporated in our 3D formalism. The comparison with numerical results in both, novel 3D and standard PW schemes, shows that non PW calculations avoid the very involved angular momentum algebra occurring for the permutations and transformations and it is more efficient and less cumbersome for considering the 3NF.

\end{abstract}

\maketitle

\section{Introduction}

As already known, the non-relativistic calculations of
few-nucleon bound and scattering states
are not able to reproduce experimental results for relevant observables
such as the triton binding, even when considering various available realistic
nucleon-nucleon (NN) interactions.
Deviations can be attributed to wrong off-shell behavior of such potentials,
 relativistic corrections and, probably more important, 3NFs effect.
In order to incorporate the 3NF corrections in a spin-isospin dependent 3D
approach \cite{Hadizadeh-MPLA24} for the triton, we have recently formulated
the corresponding Faddeev equations
in terms of the vector Jacobi momenta, specifically the magnitudes of the momenta
and the angle between them, as well as the spin-isospin quantum numbers.

We have shown that, for the full solution of the 3N bound system, the Tucson-Melbourne two-pion exchange 3NF can be included in a very simple manner in comparison to the PW representation.
As indicated in Ref. \cite{Bayegan-PRC77}, according to the number of spin-isospin states that
one takes into account, the formalism for both $^{3}H$ and $^{3}He$ bound states leads to only
strictly finite number of coupled three dimensional integral equations, which at most for fully
charge dependent case leads to 24 coupled equations.
In this communication we present our numerical results for triton binding energy, obtained by solving spin-isospin dependent three-dimensional Faddeev integral equations, when considering the Bonn-B and Tucson-Melbourne NN and 3N forces. We would like to mention that our next task, which is currently underway, is to incorporate relativistic effects in 3N
bound state calculations using the same 3D approach.

\section{Faddeev Equations in 3D representation with 3NFs }

\begin{figure}[htb]
  \includegraphics[height=.15\textheight]{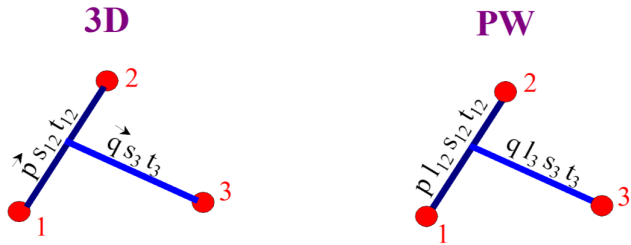}
  \caption{3N basis states in both 3D and PW representation.}
\label{fig.basis}
\end{figure}

In this section we briefly review the formalism of three-dimensional Faddeev integral equations in the realistic 3D approach. By considering the 3NF, the 3N bound state is described by the Faddeev equations
\begin{eqnarray}
| \psi \rangle = G_{0}tP |\psi\rangle +(1+G_{0}t)G_{0} V_{123}^{(3)} |\Psi\rangle,
 \label{eq.Faddeev}
\end{eqnarray}
where $G_{0}=(E-H_{0})^{-1}$ is the free 3N propagator, the operator $t=v+vG_0t$ is the NN transition matrix, $P=P_{12}P_{23}+P_{13}P_{23}$ is permutation operator, the quantity
$V_{123}^{(3)}$ defines the 3NF and $|\Psi\rangle=(1+P)|\psi\rangle$ is the total wave function.
The representation of Eq. (\ref{eq.Faddeev}) in momentum space and in a non-PW scheme needs the following
states in the 3D basis:
\begin{eqnarray}
 \quad |\, {\bf p} \, {\bf q} \,\,  \alpha  \, \rangle \equiv |  {\bf p} \, {\bf q} \,\, \alpha_{S} \,\, \alpha_{T}  \,  \rangle
 \equiv
\biggl |  {\bf p} \, {\bf q} \,\,  \left(s_{12} \,\, \frac{1}{2}\right) S \, M_{S} \,\, \left(t_{12} \,\,  \frac{1}{2}\right) T \, M_{T}  \,   \biggr  \rangle.
\label{eq.basis-3D}
\end{eqnarray}
As shown in Fig. (\ref{fig.basis}) the states of the 3D basis involve two standard Jacobi momentum vectors ${\bf p}$ and ${\bf q}$. A comparison of basis states in both 3D and PW schemes show that in a standard PW representation the angular dependence leads to two orbital angular momentum
quantum numbers, i.e., $l_{12}$ and $l_{3}$:
\begin{eqnarray}
 \quad |\, p \, q \,\,  \alpha  \, \rangle _{PW} \equiv | p \, q \,\, \alpha_{J} \,\, \alpha_{T}  \,  \rangle
\equiv   \biggl | p \, q \,\,  \biggl( (l_{12}\,\,s_{12})j_{12} \,\, (l_3\,\,\frac{1}{2}) j_3 \biggr) J \, M_{J} \,\, (t_{12} \,\,  \frac{1}{2}) T \, M_{T} \,  \biggr \rangle,
 \label{eq.basis-PW}
\end{eqnarray}
whereas in 3D representation the angular dependence explicitly appears in the Jacobi vector variables.
So it is clear that in 3D formalism there is not any coupling
between the orbital angular momenta and corresponding spin quantum
numbers. Therefore the spin quantum number of two-nucleon subsystem $s_{12}$ and the third nucleon $s_{3}=\frac{1}{2}$ couple to the total spin $S$ and its third component $M_{S}$. For the isospin quantum numbers, a similar
coupling scheme leads to the total isospin $T$ with its third component $M_{T}$.

We would like to add the remark that in order to be able to evaluate the transition and the permutation operators
we need the free 3N basis states $|\, {\bf p}\,{\bf q} \,\, \gamma
\, \rangle \equiv |\, {\bf p}\,{\bf q} \,\, m_{s_{1}} \, m_{s_{2}} \,
 m_{s_{3}} \,\, m_{t_{1}} \, m_{t_{2}} \,  m_{t_{3}}
\, \rangle $.
To this aim when we are changing the 3N basis states $| \, \alpha
\, \rangle$ to the free 3N basis states $| \, \gamma \, \rangle$
we need to calculate the Clebsch-Gordan coefficients
$ \langle \, \gamma  | \, \alpha \, \rangle \equiv g_{\gamma \alpha }
 = \langle
\, m_{s_{1}} \, m_{s_{2}} \, m_{s_{3}}  | \, (s_{12} \,\,
\frac{1}{2}) S \, M_{S} \, \rangle \, \langle \, m_{t_{1}} \,
m_{t_{2}} \, m_{t_{3}}  | \, (t_{12} \,\, \frac{1}{2}) T \, M_{T}
\, \rangle $.

By considering the symmetry property of the 3NF, the anti-symmetry property of the total wave function and the definition of anti-symmetrized NN $t$-matrix, i.e. $ _{a}\langle |t| \rangle _{a} = \langle |t(1-P_{12})|\rangle$, the representation of Eq. (\ref{eq.Faddeev}) in
Eq. (\ref{eq.basis-3D}) leads to:
\begin{eqnarray}
\label{FCs} \langle \, {\bf p}\, {\bf q}\, \alpha \, |\psi\rangle
&=& \frac{1}{{E-\frac{p^{2}}{m} -\frac{3q^{2}}{4m}}} \nonumber
\\* && \hspace{-23mm} \times \Biggl [\,
  \int d^{3}q' \, \sum_{\gamma',\gamma''',\alpha''} \,
g_{\alpha \gamma'''} \, g_{\gamma' \alpha''} \, \delta_{m'''_{s_{3}}
m'_{s_{1}}} \, \delta_{m'''_{t_{3}} m'_{t_{1}}} \,  \langle{\bf q}+\frac{1}{2}
{\bf q}' \,\, {\bf q}' \, \alpha''|\psi\rangle \nonumber
\\*  && \hspace{-21mm} \quad \times \,\, _{a}\langle{\bf p}\, m'''_{s_{1}}
m'''_{s_{2}} \, m'''_{t_{1}} m'''_{t_{2}} |t(\epsilon)
|\frac{-1}{2}{\bf q} -{\bf q}' \, m'_{s_{2}} m'_{s_{3}} \,
m'_{t_{2}} m'_{t_{3}} \rangle_{a}
 \nonumber \\*  && \hspace{-23mm} \quad +
 \Biggl\{\,\, {\langle \, {\bf p}\, {\bf q}\, \alpha \, |V_{123}^{(3)}
|\Psi\rangle}  + \frac{1}{2}
\sum_{\gamma',\gamma'',\alpha'''} \,  g_{\alpha \gamma'} \,
g_{\gamma'' \alpha'''}  \int d^{3}p'\, \frac{\delta_{m'_{s_{3}}
m''_{s_{3}}} \delta_{m'_{t_{3}} m''_{t_{3}}} }{E-\frac{p'^{2}}{m}
-\frac{3q^{2}}{4m} } \nonumber \\ && \hspace{-13mm} \times \,
_{a}\langle{\bf p}\, m'_{s_{1}} m'_{s_{2}} \, m'_{t_{1}} m'_{t_{2}}
|t(\epsilon) |{\bf p}' \, m''_{s_{1}} m''_{s_{2}} \, m''_{t_{1}}
m''_{t_{1}} \rangle_{a} \, {\langle{\bf p}'\,{\bf q} \,
\alpha'''|V_{123}^{(3)} |\Psi\rangle } \Biggr\} \, \, \Biggr].
\label{eq.Faddeev-3D}
\end{eqnarray}

As shown in Ref. \cite{Bayegan-PTP120}, the evaluation of 3NF matrix elements $\langle \, {\bf p}\,\, {\bf q} \alpha \, |V_{123}^{(3)}
|\Psi\rangle$ for $2\pi$-exchange TM force avoids the cumbersome nature of the PW representation and leads to simple expressions which are more convenient for numerical calculations.

\section{Numerical Results for $^{3}$H binding energy}

In this section we present our numerical results for the triton binding energy, obtained by solving the three-dimensional Faddeev integral equations
(\ref{eq.Faddeev-3D}). These coupled integral equations have been solved before in Ref. \cite{Bayegan-PRC77} without 3NF term.
By solving eight coupled Faddeev equations for $(\frac{1}{2}-\frac{1}{2})$ spin-isospin states and by using the operator form of Bonn-B NN potential \cite{Fachruddin-PRC62}, our calculation for the triton binding energy
converges to a value of $E_{t}=-8.152$ MeV, whereas the PW
calculations converges to $E_{t}=-8.14$ MeV for $j_{12}^{max}=4$.
In order to be able to compare our numerical results with available PW calculations, we have used TM force \cite{Coon-PRC23} (with cutoff mass $\Lambda_{\pi}=5.828 \, m_{\pi}$) in an operator form \cite{Bayegan-PTP120} which is compatible with our 3D formalism.
In table \ref{table:BE} we have shown the convergence of the triton
binding energy in 3D approach as a function of the number of grid points. The corresponding PW results are also listed as function of total angular momentum $j_{12}^{max}$. As demonstrated in this table,
by using the Bonn-B and TM combination in the 3D approach
the calculation of triton binding energy converges to $E_{t}=-9.75$ MeV,
whereas the corresponding result in a PW scheme for $j_{12}^{max}=1$ yields $E_{t}=-9.80$ MeV. We should mention that this calculation is the first attempt toward the planned numerical investigations of $^{3}H$ binding energy with the
most modern NN and 3N forces, i.e., AV18 and modified Tucson-Melbourne three-nucleon force (TM'), and also consistent NN and 3N chiral forces.

\begin{table}[hbt]
\caption {The calculated triton binding energies $E_{t}$ of the
three-dimensional Faddeev integral equations as function of the
number of grid points in Jacobi momenta $N_{jac}$ and spherical
angles $N_{sph}$, the number of grid points in polar angles is
twenty. The corresponding PW results are listed for a comparison to our results.
The used value for cutoff mass in TM 3NF is $\Lambda_{\pi}=5.828 \, m_{\pi}$.}
\begin{tabular}{lcccccccccccl}
\hline
Potential &&& &&& &&& &&& $E_{t}$ [MeV] \\
\hline
\multicolumn{13}{c}{3D approach}\\
\hline
Bonn-B \\
&&& &&& $N_{jac}$  &&&  $N_{sph}$ &&&   \\
&&& &&& 40  &&&  24 &&& -8.15 \cite{Bayegan-PRC77} \\
Bonn-B + TM \\
&&& &&& 32  &&&  20 &&& -9.73  \\
&&& &&& 32  &&&  24 &&& -9.73  \\
&&& &&& 36  &&&  20 &&& -9.74  \\
&&& &&& 36  &&&  24 &&& -9.74  \\
&&& &&& 40  &&&  20 &&& -9.75  \\
&&& &&& 40  &&&  24 &&& -9.75  \\
&&& &&& 40  &&&  32 &&& -9.75  \\
\hline
\multicolumn{13}{c}{PW approach}\\
\hline
Bonn-B
\\
 &&&  &&&  &&&  $j_{12}^{max}$ \\
 &&&  &&&  &&& 1 &&& -8.17 \cite{Gloeckle-NPA560}    \\
 &&&  &&&  &&& 2 &&& -8.10 \cite{Schadow-FBS28} \\
 &&&  &&&  &&& 3 &&& -8.14 \cite{Gloeckle-PRL71}  \\
 &&&  &&&  &&& 4 &&& -8.14 \cite{Sammarruca-PRC46} \\
\\
Bonn-B + TM  &&& &&& &&& 1   &&& -9.80 \cite{Gloeckle-NPA560}\\
\hline
\end{tabular}
\label{table:BE}
\end{table}

\begin{theacknowledgments}
M. R. Hadizadeh and L. Tomio would like to thank the Brazilian agencies FAPESP and CNPq for partial support. S. Bayegan acknowledges the support of center of excellence on structure of matter, Department of Physics, University of Tehran.
\end{theacknowledgments}

\bibliographystyle{aipproc}   

\end{document}